\RequirePackage{fix-cm}
\documentclass[smallextended]{svjour3}       
\smartqed  
\usepackage{amsmath}
\usepackage{mathtools}
\usepackage{graphicx}
\usepackage{lscape}              
\usepackage{footnote}
\usepackage{verbatim}
\usepackage{epstopdf}
\usepackage{graphics}
\usepackage{bm}
\usepackage{amsmath}
\newcommand{\squeezeup}{\vspace{-3.0mm}}
\begin{document}
\title{Antiproton low-energy collisions with Ps-atoms and true muonium
atoms\hspace{0.886mm}($\mu^+\mu^-$)}

\subtitle{$\overline{\rm{H}}$ and $\overline{\rm{H}}_{\mu}$ three-body formation reactions}


\author{Renat A. Sultanov         \and
        D. Guster 
}


\institute{Renat A. Sultanov \and D. Guster$^{\dagger}$ \at
              Department of Information Systems \&  
              Integrated Science and Engineering Laboratory Facility {\it(ISELF)}
              at St. Cloud State University, St. Cloud, MN 56301-4498, USA \\
              Tel.: +1-308-5756\\
              Fax: +1-308-6074\\
              \email{rasultanov@stcloudstate.edu} \\
              \email{dcguster@stcloudstate.edu}$^{\dagger}$
}
\date{Received: date / Accepted: date}
\maketitle

\begin{abstract}
Three-charge-particle collisions with participation of ultra-slow
antiprotons ($\overline{\rm{p}}$) is the subject of this work.
Specifically we compute the total cross sections and corresponding thermal rates
of the following three-body reactions:
$\overline{\rm p}+(e^+e^-) \rightarrow \overline{\rm{H}} + e^-$ and
$\overline{\rm p}+(\mu^+\mu^-) \rightarrow \overline{\rm{H}}_{\mu} + \mu^-$, where
$e^-(\mu^-)$ is an electron (muon) and $e^+(\mu^+)$
is a positron (antimuon) respectively,
$\overline{\rm{H}}=(\overline{\rm p}e^+)$ is an antihydrogen atom and
$\overline{\rm{H}}_{\mu}=(\overline{\rm p}\mu^+)$
is a muonic antihydrogen atom, i.e. a bound state of $\overline{\rm{p}}$ and $\mu^+$.
A set of two-coupled few-body Faddeev-Hahn-type (FH-type) equations is numerically solved
in the framework of a modified close-coupling expansion approach.
\end{abstract}

\squeezeup
\section{Introduction}
\label{intro}
Recently created ultra-low energy antiprotons are of great scientific interest because of the possible
formation of slow antihydrogen atoms \cite{gabrielse2011,andresen2010}.
The main motivation of the antihydrogen and antimatter physics research
is to check and confirm (or not confirm) certain fundamental laws and theories of modern physics.
For example, one of the most important subjects in the field is to check the charge conjugation,
parity, and time reversal (CPT) symmetry of quantum electrodynamics. In other words:
a charged particle and its antiparticle should have equal/opposite charges, equal masses, lifetimes, 
and gyromagnetic ratios.
In quantum field theory the CPT theorem plays a very important and fundamental role.
It states that any canonical (local, Lorentz-covariant) quantum field theory is invariant 
under the CPT operation.
%
The CPT symmetry also predicts that hydrogen and antihydrogen 
atoms should have identical spectra. In order to test these fundamental 
laws of physics new experiments are in progress.
It is planed to test whether H and $\overline{\rm{H}}$ have such properties.
In such sensitive experiments, it would be important to have
a certain quantity of $\overline{\rm{H}}$ atoms at low kinetic energies,
ideally at  rest: $T\sim0$ K. Next,
in this context it would be useful to mention muonic
physics problems and related muonic-atomic few-body systems too. In this field
a scientific breakthrough has recently been achieved: it was found that the size of a proton in the
muonic hydrogen atom $({\rm{p}}\mu^+)$
differs by $\sim$4\% from the size of a proton in normal hydrogen atom \cite{pohl}. It is clear, that
it would be extremely interesting to undertake
the same comparisons between antihydrogen and muonic antihydrogen atoms too.
The author of \cite{nagam03a} pointed out that the 
muonic antihydrogen atom, $\overline{\rm{H}}_{\mu}$,
which is a bound state of $\overline{\rm{p}}$ and an antimuon, $\mu^+$,
could be an even better choice to check the CPT law than the usual
antihydrogen atom. This is because the size of this atom is $\sim$ 207 times smaller
than the size of a normal $\overline{\rm{H}}$ atom. Therefore, as mentioned in \cite{nagam03a}:
the short range CPT violating interaction with an extremely heavy boson can be easily detected within the system.
Further, in a relatively old work \cite{ponomarev} the authors considered the reaction of $\mu$-capture by
hydrogen nuclei, i.e. the following process:
\begin{equation}
\mu^- + {\rm p}\rightarrow {\rm n} + \nu.
\label{eq:capture}
\end{equation}
The study of the capture of negative muons by atomic nuclei can provide valuable information about
the weak interaction \cite{primakoff} and the reaction (\ref{eq:capture}) can be used to determine the
weak interaction constants \cite{ponomarev}. One of the most important characteristics of the process
is the rate of the nuclear $\mu$-capture.
The rate of (\ref{eq:capture}) significantly depends on the mutual muon and nuclear spin orientation.
In \cite{ponomarev} this process was considered within the muonic molecular ion $(\rm{pp}\mu^-)^+$. Now, in our opinion,
it would also be very interesting to transfer this consideration to antimatter physics,
i.e. consider the antimuon $\mu^+$-capture
by antiprotons. For example, in slow collisions between $\overline{\rm{H}}$ and $\overline{\rm{H}}_{\mu}$ one can
form an antimuonic molecular ion $(\bar{\rm{p}}\bar{\rm{p}}\mu^+)^-$:
\begin{equation}
\overline{\rm{H}} + \overline{\rm{H}}_{\mu}\rightarrow (\bar{\rm{p}}\bar{\rm{p}}\mu^+)^- + e^+.
\label{eq:ppmu}
\end{equation}
In this reaction a part of the excess energy after the formation of $(\bar{\rm{p}}\bar{\rm{p}}\mu^+)^-$
is taken by free $e^+$. In conclusion,
the low energy formation of $\overline{\rm{H}}_{\mu}$ can be achieved with
the use of the true muonium atom $(\mu^+\mu^-)$:
the smallest pure QED atom with a Bohr radius only $\sim$512 fm.

These ideas appear be interesting, and therefore it would be useful to compute
the formation cross sections and rates of $\overline{\rm{H}}_{\mu}$ at low energy collisions, for example,
from $\sim$1 eV down to $\sim10^{-5}$ eV. Thus, in this work we consider the following three-body reactions
of antihydrogen $\overline{\rm{H}}$ and muonic antihydrogen $\overline{\rm{H}}_{\mu}$ formation:
\begin{equation}\label{eq:Pse}
\overline{\rm p}+(e^+ e^-)_{1s}\rightarrow\overline{\rm{H}} + e^-,
\end{equation}
\begin{equation}\label{eq:Psmu}
\overline{\rm p}+(\mu^+\mu^-)_{1s}\rightarrow\overline{\rm{H}}_{\mu} + \mu^-.
\end{equation}
At such low energies the quantum-mechanical
Coulomb few-body dynamics becomes important, especially in the case of heavy charge
transfer, i.e. $\mu^+$. Also, it would be quite appropriate to mention that
exotic atomic and antiatomic systems like a true muonium atom, $(\mu^+\mu^-)$,
or a simple muonic hydrogen atom, H$_{\mu}$=(p$^+\mu^-)$, are always of great interest in nuclear, atomic and 
few-body physics. There is also another three-charge-particle reaction of interest
in $\overline{\rm{H}}_{\mu}$ formation:
\begin{equation}
\overline{\rm p} + \mbox{Mu} \rightarrow \overline{\rm{H}}_{\mu} + e^-.
\end{equation}
Here, Mu is the muonium atom, i.e. a bound state of a positive muon (antimuon)
$\mu^+$ and an electron: Mu=$(\mu^+e^-)$.
This is a very interesting and challenging computation example of a heavy charge, $\mu^+$,
transfer reaction.

Using this perspective we develop a quantum-mechanical
approach which should be reliable at low and very low collision energies, 
i.e. when the quantum-mechanical few-body dynamics of three Coulomb particles
becomes important. The method is formulated for arbitrary masses of the particles, that is
when the dynamics of lighter and heavier particles are not separated from each other.
In the current work we apply a {\it few-body} approach based on a set of coupled two-component
FH-type equation formalism \cite{my_arxiv,my_fbs}.


\squeezeup
\section{Basic equations and results}
\label{sec:1}
Let us define the system of units to be $e=\hbar=m_3=1$ and
denote antiproton $\overline{\rm{p}}$ by 1, a negative muon $\mu^-$  by 2, and a positive muon $\mu^+$ by 3.
Before the three-body breakup threshold two cluster asymptotic configurations
are possible in the three-body system, i.e. (23)$-$1 and
(13)$-$2  being determined by their own Jacobi coordinates $\{\vec r_{j3}, \vec \rho_k\}$:
\begin{eqnarray}\label{eq:coord}
\vec r_{j3} = \vec r_3 - \vec r_j,\ \ 
\vec \rho_k = {(\vec r_3 + m_j\vec r_j)} / {(1 + m_j)} - \vec r_k,\ \ 
(j \not = k=1, 2).
\end{eqnarray}
Here $\vec r_{\xi}$, $m_{\xi}$ are the coordinates and the
masses of the particles $\xi=1, 2, 3$ respectively.
This suggests a Faddeev formulation which uses only two components.
In this approach the three-body wave function is represented as follows:
\begin{equation}\label{eq:psi3b}
|\Psi\rangle =  \Psi_1 (\vec r_{23},\vec \rho_1) + \Psi_2 (\vec r_{13},\vec \rho_2),
\end{equation}
where each Faddeev-type component is determined by its own Jacobi coordinates. Moreover,
$ \Psi_1 (\vec r_{23}, \vec \rho_1)$ is quadratically integrable
over the variable $\vec r_{23}$, and $\Psi_2 (\vec r_{13},\vec
\rho_2)$ over the variable $\vec r_{13}$. To define $|\Psi_l\rangle$, $(l = 1, 2)$ a set of two coupled
FH-type equations can be written:
\begin{equation}\label{eq:fh1}
\Big (E-\hat{H}_0-V_{23}(\vec r_{23})  
\Big ) \Psi_1 (\vec r_{23}, \vec \rho_1) =  
\Big (V_{23}(\vec r_{23}) + V_{12}(\vec r_{12}) 
\Big )\Psi_2 (\vec r_{13}, \vec \rho_2),
\end{equation}
\begin{equation}\label{eq:fh2}
\Big (E-\hat{H}_0-V_{13}(\vec r_{13}) 
\Big )\Psi_2 (\vec r_{13}, \vec \rho_2) =  
\Big (V_{13} (\vec r_{13}) + V_{12}(\vec r_{12}) 
\Big )\Psi_1 (\vec r_{23}, \vec \rho_1).
\end{equation}
Here, $\hat{H}_0$ is the kinetic energy operator of the
three-particle system, $V_{ij} (r_{ij})$
are paired interaction potentials $(i \not= j = 1,2,3)$,
$E$ is the total energy.
%
\begin{table}\label{tab2}
\caption{The total cross sections $\sigma_{\overline{\rm H}}$ and 
$\sigma_{\overline{\rm H}_{\mu}}$ for the reactions (\ref{eq:Pse}) and  (\ref{eq:Psmu}) respectively.
The product of these cross sections and the corresponding center-of-mass velocities $v_{c.m.}$
between $\overline{\rm{p}}$ and Ps=$(e^+ e^-)$, i.e. $\sigma_{\overline{\rm H}} v_{c.m.}$
and between $\overline{\rm{p}}$ and the true muonium atom Ps$_{\mu}=(\mu^+ \mu^-)$, i.e. 
$\sigma_{\overline{\rm H}_{\mu}} v_{c.m.}$ are presented.}
\vspace{2mm}
\centering
\begin{tabular}{lcccccc}
\hline\noalign{\smallskip}
&\multicolumn{2}{c}{$\overline{\rm p}+(e^+ e^-)_{1s}\rightarrow\overline{\rm{H}} + e^-$} &
&\multicolumn{2}{c}{$\overline{\rm p}+(\mu^+\mu^-)_{1s}\rightarrow\overline{\rm{H}}_{\mu} + \mu^-$}\\
\hline\noalign{\smallskip}
$E$, eV & $\sigma_{\overline{\rm H}}$,\ cm$^2$
              & $\sigma_{\overline{\rm H}}  v_{c.m.}$,\ cm$^3$/s &
              & $\sigma_{\overline{\rm H}_{\mu}}$,\ cm$^2$
              & $\sigma_{\overline{\rm H}_{\mu}}  v_{c.m.}$,\ cm$^3$/s\\
\hline\noalign{\smallskip}
1.0E-06 & 0.16E-12 & 0.67E-08 &      &                  &               \\
1.0E-05 & 0.50E-13 & 0.67E-08 &      &                  &               \\
1.0E-04 & 0.16E-13 & 0.67E-08 &      &  0.18E-16 & 0.60E-12\\ 
1.0E-03 & 0.50E-14 & 0.66E-08 &      &  0.58E-17 & 0.60E-12\\ 
1.0E-02 & 0.15E-14 & 0.63E-08 &      &  0.18E-17 & 0.59E-12\\ 
5.0E-02 & 0.60E-15 & 0.56E-08 &      &  0.82E-18 & 0.59E-12\\
1.0E-01 & 0.42E-15 & 0.55E-08 &      &  0.58E-18 & 0.59E-12\\
5.0E-01 &                &                 &      &  0.27E-18 & 0.62E-12\\
1.0E-00 &                &                 &      &  0.23E-18 & 0.73E-12\\
\hline\noalign{\smallskip}
\end{tabular}
\end{table}
\begin{figure*}
\begin{center}
%
%
\includegraphics[height=0.37\textwidth,width=0.47\textwidth]{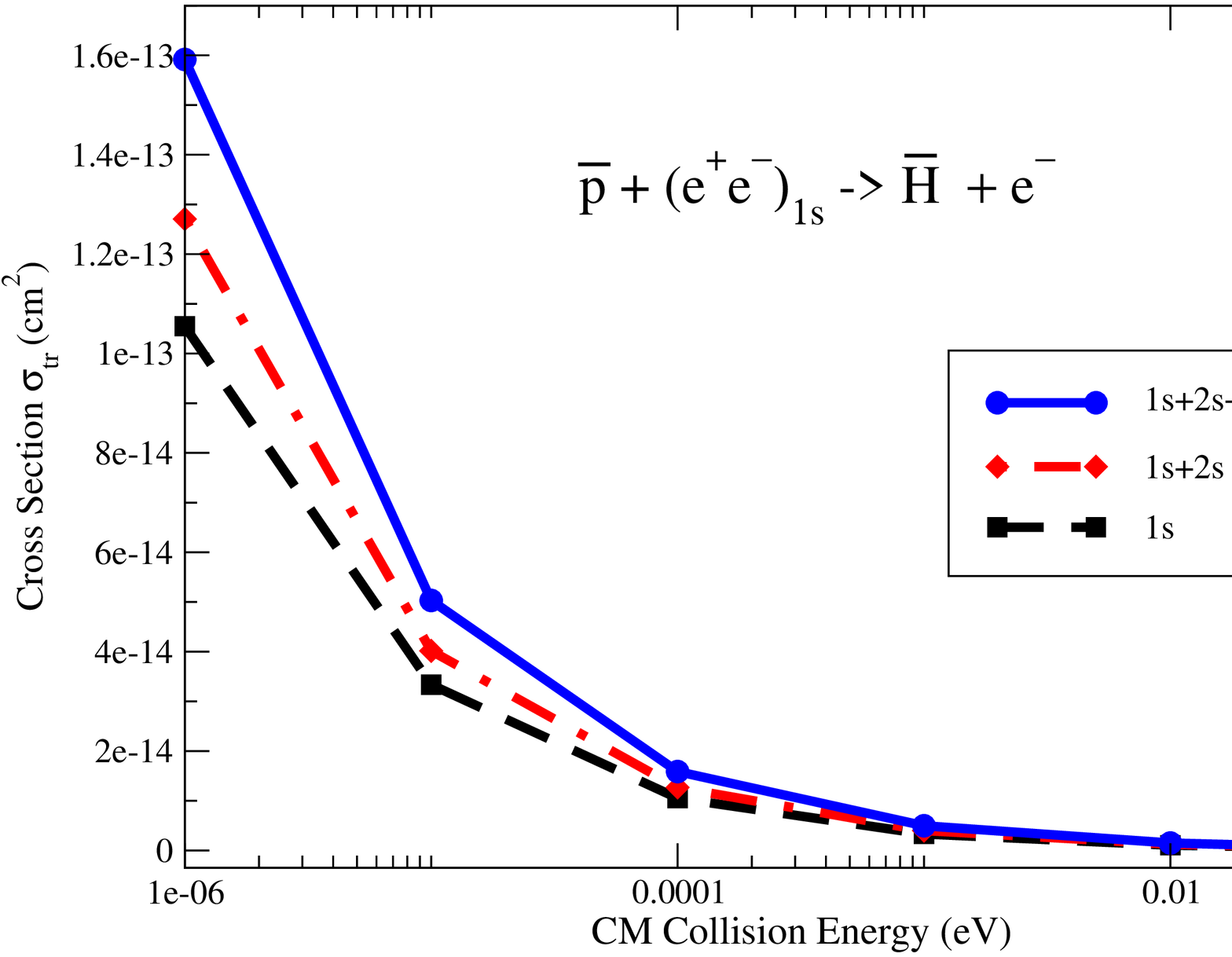}
\vspace{1mm}\\
\includegraphics[height=0.37\textwidth,width=0.47\textwidth]{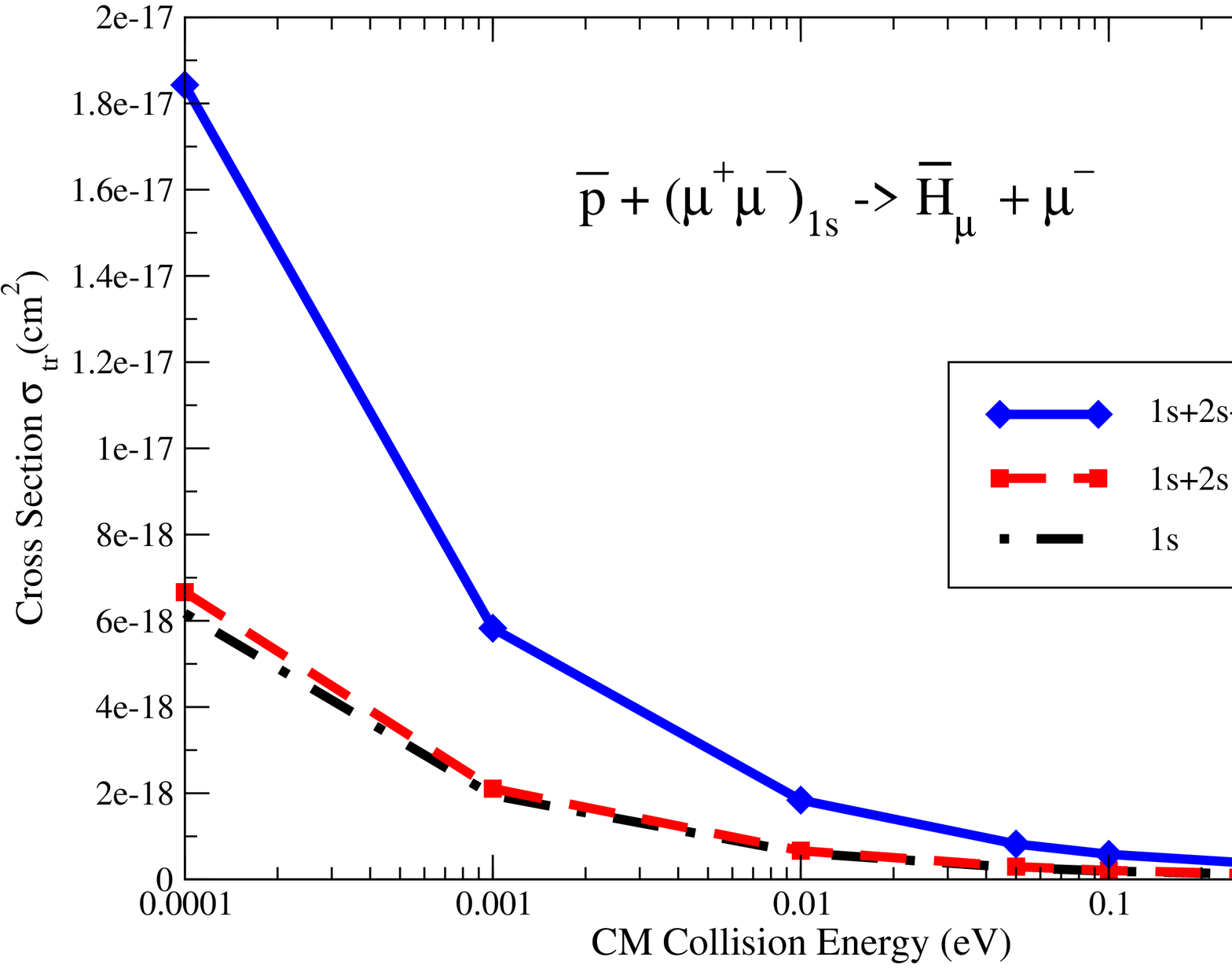}
\vspace{1mm}\\
\includegraphics[height=0.37\textwidth,width=0.47\textwidth]{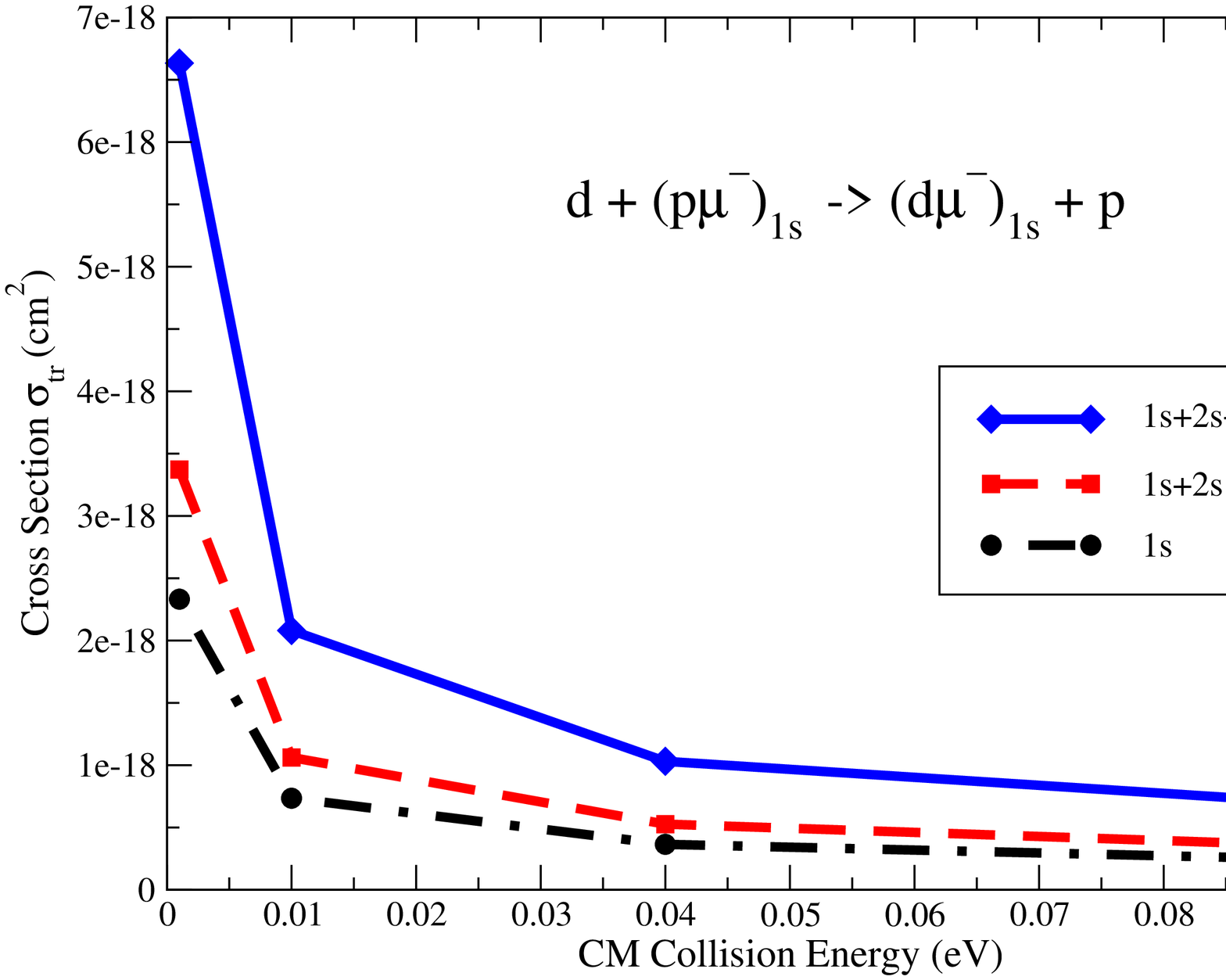}
\vspace{5mm}\\
%
\caption{Low energy cross sections of ${\overline{\rm H}}$ (upper plot)
and ${\overline{\rm H}_\mu}$ (middle plot) formation reactions
are shown. On the lower plot the cross section for the $\mu^-$ transfer from a proton to a deuteron is also
included as a test. This result provides the muon transfer thermal rate close to the experimental data
at low temperatures \cite{my_arxiv}. The cross sections are computed within the two-level 2$\times$1s,
four-level 2$\times$(1s+2s), and  six-level 2$\times$(1s+2s+2p) approximations.}
\end{center}
\label{fig:2}       
\end{figure*}
The constructed equations satisfy the Schr\H{o}dinger
equation exactly. For the energies below the three-body
break-up threshold these equations exhibit the same advantages as the
Faddeev equations,  
because they are formulated for the wave
function components with the correct physical asymptotes.
To solve the  equations a modified close-coupling
method is applied, which leads to an expansion
of the system's wave function components into
eigenfunctions ($\varphi^{(i)}_{n(n')}(\vec r_{j3}),\ i\not=j=1,2$) of the subsystem (target) Hamiltonians, i.e.
\setlength{\textfloatsep}{2pt}
\begin{equation}
\Psi_1(\vec r_{23}, \vec \rho_1)\approx\ {\large{{\mathclap{\displaystyle\int}\mathclap{\sum_{n}}}}}
\hspace{3mm}f^{(1)}_n(\vec \rho_1)\varphi^{(1)}_n(\vec r_{23}),\ \ 
\Psi_2(\vec r_{13}, \vec \rho_2)\approx \ {\large{{\mathclap{\displaystyle\int}\mathclap{\sum_{n'}}}}}
\hspace{3mm}f^{(2)}_{n'}(\vec \rho_2)\varphi^{(2)}_{n'}(\vec r_{13}).
\label{eq:fh77rs}
\end{equation}
It provides us with a set of one-dimensional integral-differential equations
after the partial-wave projection. A further advantage of the Faddeev-type
method is the fact that the Faddeev-components are smoother functions of
the coordinates than the total wave function.
Also, the Faddeev decomposition avoids overcompleteness problems, because
two-body subsystems are treated in an equivalent way, and the correct
asymptotes are guaranteed.
After a proper angular momentum expansion one can obtain an infinite set of
coupled integral-differential equations for the unknown functions $f_{\alpha}^{(1)}(\rho_1)$ and
$f_{\alpha^\prime}^{(2)}(\rho_2)$ \cite{my_arxiv,my_fbs}:
\begin{eqnarray}
\Big( (k^i_n)^2 + \frac{\partial^2}
{\partial \rho_i^2} -
\frac{\lambda (\lambda + 1)}{\rho_i^2}
\Big) f_{\alpha}^{(i)}(\rho_i)\ =\ g \sum_{\alpha'} 
\sqrt{\frac{(2\lambda + 1)(2\lambda^{\prime} + 1)}  
{(2L+1)}}\nonumber \\
 \int_{0}^{\infty} d \rho_{i'}
f_{\alpha^\prime}^{(i')}(\rho_{i'})
\int_{0}^{\pi}
d \omega \sin\omega
R_{nl}^i(r_{i'3})
\Big(V_{i'3}(r_{i'3}) + V_{ii'}(r_{ii'})\Big)\rho_{i'} \rho_i
\nonumber \\
 R_{n'l'}^{i'}(r_{i3})
\sum_{mm'} D_{mm'}^L(0, \omega, 0)
C_{\lambda 0lm}^{Lm} C_{\lambda' 0l'm'}^{Lm'}
Y_{lm}(\nu_i, \pi)Y^{*}_{l'm'}(\nu_{i'}, \pi).
\label{eq:fh7}                                                                                                        
\end{eqnarray}
\setlength{\textfloatsep}{2pt}
For the sake of simplicity $\alpha \equiv (nl\lambda)$ are quantum numbers of a
three-body state \cite{my_arxiv} and $L$ is the total angular momentum of the
three-body system, $g=4\pi M_i/\gamma^{3}$,
$k^i_n = \sqrt{2M_i(E-E_n^{i'})}$, where $E_n^{i'}$ is the binding energy of the subsystem $(i^{\prime}3)$,
$M_1=m_1(m_2+m_3)/(m_1+m_2+m_3)$ and
$M_2=m_2(m_1+m_3)/(m_1+m_2+m_3)$ are the reduced masses,
$\gamma=1-m_im_{i'}/((m_i+1)(m_{i'}+1))$,
$D_{mm'}^L(0, \omega, 0)$ the Wigner functions,
$C_{\lambda 0lm}^{Lm}$ the Clebsh-Gordon coefficients,
$Y_{lm}$ are the spherical functions,
$\omega$ is the angle between the Jacobi coordinates
$\vec \rho_i$ and $\vec \rho_{i'}$, $\nu_i$ is the angle between 
$\vec r_{i'3}$ and $\vec \rho_i$, $\nu_{i'}$ is the angle
between $\vec r_{i3}$ and $\vec \rho_{i'}$ \cite{my_arxiv,my_fbs}.
We numerically solve the set of coupled integral-differential equations
(\ref{eq:fh7}) together with specific boundary conditions which are
appropriate for the three-body rearrangement scattering problems
(\ref{eq:Pse}) and (\ref{eq:Psmu}) \cite{my_arxiv,my_fbs}.
Also, additionally we compute the $\mu^-$ transfer reaction from one hydrogen isotope to another heavier hydrogen isotope.
Now, below we briefly discuss our computational results.
All these different three-body Coulomb systems have been computed in the framework of
a {\it unique} quantum-mechanical method, i.e. the FH-type equation formalism (\ref{eq:fh1})-(\ref{eq:fh2})
and (\ref{eq:fh77rs}).
The details of the method have been presented in our recent preprint \cite{my_arxiv}.
The goal of these works is to carry out a quantum-mechanical calculation of the formation cross sections
and corresponding thermal rates 
of the $\overline{\rm{H}}$ and $\overline{\rm{H}}_{\mu}$ atoms at very
low collision energies, i.e. the reactions (\ref{eq:Pse}) and (\ref{eq:Psmu}).
The coupled integral-differential Eqs. (\ref{eq:fh7}) have been solved numerically for the case of
the total angular momentum $L=0$ within the two-level  2$\times$(1s), four-level 2$\times$(1s+2s), 
and six-level 2$\times$(1s+2s+2p) close coupling approximations in Eqs. (\ref{eq:fh77rs}). The sign "2$\times$"
indicates that two different sets of expansion functions are applied.
To compute the charge transfer cross sections a {\it K-}matrix formalism has been used \cite{my_arxiv,my_fbs}.
Table I shows our results for the cross sections $\sigma_{\overline{\rm{H}}}$ and $\sigma_{\overline{\rm{H}}_{\mu}}$
for the reactions (\ref{eq:Pse}) and (\ref{eq:Psmu}) respectively. Also, Table I represents our data for the products
$\sigma_{\overline{\rm{H}}}v_{cm}$ and $\sigma_{\overline{\rm{H}}_{\mu}}v_{cm}$, where $v_{cm}$ is relative
velocities between colliding particles. One can see, that at very low energies they take almost constant values.
These results are in good agreement with the general rule of the quantum-mechanical rearrangement
scattering theory: $\sigma_{tr}\sim 1/v_0$, where $\sigma_{tr}$ is the transfer cross-section and
$v_0$ is the velocity in the input channel. Now these quantities can
be used for actual computation of $\overline{\rm{H}}$ and $\overline{\rm{H}}_{\mu}$ production rates \cite{my_arxiv}.
Finally, Fig. 1 shows our cross sections for the antihydrogen processes (\ref{eq:Pse}) and (\ref{eq:Psmu})
and one muon transfer reaction, i.e. $\rm{d}+(\rm{p}\mu)_{1s}\rightarrow (\rm{d}\mu)_{1s}+\rm{p}$. These results
are shown within different close-coupling expansion approximations in Eqs. (\ref{eq:fh77rs}). One can see that
contribution of the polarization interaction ({\it p-}wave) becomes significant while the collision energy decreases.
Among the considered systems in this work the biggest {\it p-}wave contribution at low energies 
is identified in the interesting $\overline{\rm{p}}+(\mu^+\mu^-)$ collision. There may be different
physical reasons for this phenomenon as, for example, the mass symmetry, inertia, and the charge asymmetry in the target.
Presumably, the vacuum polarization effects(Casimir) and
Casimir-Polder-type forces may also be significant in this exotic system.

\squeezeup


\begin{thebibliography}{77}
\squeezeup
\bibitem{gabrielse2011}
Gabrielse, G., et al., (ATRAP Collaboration):
Adiabatic cooling of antiprotons.
Phys. Rev. Lett. {\bf 106}, 073002 (2011).

\bibitem{andresen2010}
Andresen, G.B. et al., (ALPHA Collaboration):
Evaporative cooling of antiprotons to cryogenic temperatures.
Phys. Rev. Lett. {\bf 105}, 013003 (2010).

\bibitem{pohl}
Pohl, R., et al.: The size of the proton.
Nature {\bf 466}, 213 (2010).

\bibitem{nagam03a}
Nagamine, K.:
Introductory muon science, Cambridge University Press, Cambridge (2003).

\bibitem{ponomarev}
Bakalov, D.D., Faifman, M.P., Ponomarev, L.I., and Vinitsky, S.I.:
$\mu$-Capture and ortho-para transitions in the muonic molecule $({\rm{p}}{\rm{p}}\mu)$.
Nucl. Phys. A {\bf 384}, 302 (1982).

\bibitem{primakoff}
Primakoff, H.:
Theory of muon capture. Rev. Mod. Phys. {\bf 31}, 802 (1959).

\bibitem{my_arxiv}
Sultanov, R.A. and Guster, D.: 
Antihydrogen $(\bar {\rm H})$ and muonic antihydrogen
$(\bar {\rm H}_{\mu})$ formation in low energy three-charge-particle collisions. arXiv: 1304.2434v2.

\bibitem{my_fbs}
Sultanov, R.A. and Guster, D.:
Muonic antihydrogen $(\bar {\rm H}_{\mu})$ formation in low-energy three-body reactions.
Few-Body Syst. {\bf 54}, 1157 (2013).

\end{thebibliography}
\end{document}